%% file: mingaliev.tex
\documentclass[
 aps,
 10pt,
 final,
 notitlepage,
 oneside,
 onecolumn,
 nobibnotes,
 nofootinbib,
 superscriptaddress,
 noshowpacs,
 centertags]
{revtex4}

\input{sao_cmd_author.tex}

\usepackage{textcomp}
\usepackage{xcolor}
\usepackage{comment}
\usepackage{url}
\usepackage{multirow}
\usepackage{longtable}

\begin{document}

\title{The synchrotron component study in a spectral energy distribution of blazars}

\author{\firstname{M.~G.}~\surname{Mingaliev}}
\affiliation{Special Astrophysical Observatory of RAS, Nizhnij Arkhyz, 369167 Russia}
\affiliation{Kazan Federal University, 18 Kremlyovskaya St., Kazan, 420008, Russia}
\email[E-mail:]{marat@sao.ru}

\author{\firstname{Yu.~V.}~\surname{Sotnikova}}
\affiliation{Special Astrophysical Observatory of RAS, Nizhnij Arkhyz, 369167 Russia}

\author{\firstname{T.~V.}~\surname{Mufakharov}}
\affiliation{Special Astrophysical Observatory of RAS, Nizhnij Arkhyz, 369167 Russia}
\email{timur.mufakharov@gmail.com}

\author{\firstname{A.~K.}~\surname{Erkenov}}
\affiliation{Special Astrophysical Observatory of RAS, Nizhnij Arkhyz, 369167 Russia}

\author{\firstname{R.~Yu.}~\surname{Udovitskiy}}
\affiliation{Special Astrophysical Observatory of RAS, Nizhnij Arkhyz, 369167 Russia}



\begin{abstract}
We study the synchrotron component of the spectral energy distribution (SED)
on the sample of 877 blazars using the
ASDC SED Builder Tool with available broadband data from the literature.
Our sample includes 423 flat-spectrum radio sources (FSRQs),
361 BL Lac objects and candidates, and 93 blazars of uncertain type.
We have made an estimation of the synchrotron peak frequency ($\nu_{peak}^{s}$)
for the 875 objects and further classified them as
high, intermediate and low synchrotron peaked sources (HSPs/ISPs/LSPs).
There are 42 HSPs with $\nu_{peak}^{s} > 10^{16.5}$ Hz,
222 ISPs with $10^{14.5} < \nu_{peak}^{s} < 10^{16.5}$ Hz,
and 611 LSPs with $\nu_{peak}^{s} < 10^{14.5}$ Hz
in our sample.
We have calculated an average value of $\nu_{peak}^{s}$ to be
$10^{13.4 \pm 1.0}$ Hz for FSRQs and $10^{14.6 \pm 1.4}$ Hz for BL Lacs.
We found out that $\nu_{peak}^{s}$ and
the flux density at 4.8 GHz have a different distribution
(as indicated by Kolmogorov--Smirnov test at significance level 0.05)
for the FSRQ and BL Lac blazars, and for the RBL and XBL types of BL Lacs.
Distribution of $\nu_{peak}^{s}$ values is broader for BL Lacs,
than for FSRQs.
There are no ultra-high energy peaked objects (with $\nu_{peak}^{s} > 10^{19}$ Hz)
in our BL Lac sample according to our estimations.
The significant part of FSRQs (41\%) and small part of BL Lacs (9\%) in our sample
could be considered as candidates to the very-low synchrotron peaked
blazars (with $\nu_{peak}^{s} < 10^{13}$ Hz).
Our foundations confirm results of the previous studies made on
 samples with significantly smaller number of objects.
\end{abstract}

\maketitle
{\small Keywords: {quasars: general---BL Lacertae objects: general---galaxies: nuclei---galaxies: jets---
radio continuum: galaxies}\par}

\textit{Will appear in Astrophysical Bulletin, Volume 70, Issue 3, pp.264-272, 2015}

\section{Introduction}
The blazars are active galactic nuclei (AGNs) with the jet, viewed at small angles to the observer~\cite{1995PASP..107..803U}.
This characteristic feature explains various
observational properties of this class of objects.
Due to strong magnetic fields in the jet, non-thermal
radiation dominates over the entire range of the electromagnetic
spectrum of blazars. Blazars are historically
divided into two subclasses: flat-spectrum radio
quasars (FSRQs) and BL Lacertae type objects (BL Lacs). The optical
spectrum of FSRQs reveals strong broad emission
lines, while the spectrum of BL Lac objects often
have no lines, sometimes possessing weak emission
or absorption lines. Blazars are characterized by variable
non-thermal radiation in all frequency bands.
The spectral energy distribution (SED)
of blazars has two characteristic components: a low frequency component
with a peak in the optical, ultraviolet, or X-ray
spectral region and a high frequency component with
a peak in the gamma-ray band. Their presence is
usually explained by synchrotron radiation and the
effect of inverse Compton scattering~\cite{1996ApJ...463..444S}.
Most of the blazar radiation from the radio to optical
band (and in some cases, in X-rays) is the synchrotron
radiation of charged particles in the jet (see,
e.g., \cite{1981Natur.293..714B,1982ApJ...253...38U,
1988AJ.....95..307I,1998ASPC..144...25M}).

In addition to the division of blazars into the
FSRQ and BL Lac types by the presence or absence
of lines in the optical spectrum, there is also a classification
of blazars by the synchrotron peak frequency ($\nu_{peak}^{s}$) in their SED.
Blazars divided into high/intermediate/low-synchrotron peaked, according to the $\nu_{peak}^{s}$:
high synchrotron peaked~(HSPs) ones with the $\nu_{peak}^{s}>10^{16.5}$ Hz,
 low synchrotron peaked~(LSPs) blazars with the $\nu_{peak}^{s}<10^{14.5}$ Hz,
 and intermediate synchrotron-peaked (ISPs) have $10^{14.5}<\nu_{peak}^{s}<10^{16.5}$ Hz. 
In this paper we use this classification, proposed by Urry and Padovani~\cite{1995PASP..107..803U}.

The BL Lac blazars are traditionally divided into RBLs (radio-selected), OBLs
(optical-selected) and XBLs (X-ray-selected) depending
on the band in which they were originally identified: radio, optical, or X-rays
\cite{1991ApJ...374..431S,1994A&AS..103..349S,1990AJ.....99....1K,
1984ApJ...284L..23M,1990ApJS...72..567G,1985ApJ...298..619S,1996ApJS..104..251P}.
Typically, they differ in the position of the synchrotron component
on the SED. For the RBLs, the synchrotron
component peak more often falls into the radio to IR
frequency range, for the XBL-objects it resides
in the UV-X-ray band \cite{1995A&AS..109..267G}.
Such a historical division
of BL Lac objects into XBLs and RBLs is often not
related to the physical differences of the objects themselves \cite{1995PASP..107..803U}.
Some BL Lacs that were not detected
in the X-rays are regarded as XBL blazars due to
the high X-ray/radio flux ratio \cite{1994MNRAS.268L..51G,1994PhDT........43W}.
For example, PKS\,0548$-$32, PKS\,2005$-$48 and
PKS\,2155$-$30, or optically identified Mrk\,180, Mrk\,421 and Mrk\,501
are considered as XBL-type \cite{2013AJ....145...31K}.

The shape of the SEDs of the blazars
and the $\nu_{peak}^{s}$ value may
change depending on the activity of the object, sometimes
significantly, by orders of magnitude \cite{2014MNRAS.445.4316C,2011A&A...529A.145D}.
In those cases object could be considered as a ``transition'' type,
within the simplified model in which the blazar
type and its radio luminosity are determined by the
degree of activity of a given radio galaxy
(FRI and FRII) \cite{2012MNRAS.420.2899G,2013MNRAS.431.1914G}.

Nieppola, Tornikoski, and Valtaoja \cite{2006A&A...445..441N} determined 
 $\nu_{peak}^{s}$ for about 300 BL~Lacs using the literature data. Subsequently, researchers of
broadband properties of blazars referred
to this study, even though the literature data it used
was non-homogeneous (except for the radio band
data). Over the recent years, the number of blazars
with available observed data has greatly increased.
At the same time, there appeared new measurements
for the objects from this sample, defining the
position of $\nu_{peak}^{s}$. In this study, we list the 
ultra-high-energy synchrotron-peak (UHBLs)
BL Lac candidates, with $\nu_{peak}^{s} > 10^{19}$Hz.
For half of them, according to our estimates, $\nu_{peak}^{s}$ values
appeared to be significantly smaller, although still
almost all of these objects can be referred to as
HSP ($\nu_{peak}^{s} > 10^{16.5}$Hz). In \cite{2008A&A...488..867N} the same group of
researchers determined $\nu_{peak}^{s}$ for 135 blazars.

Giommi et al. \cite{2012A&A...541A.160G} have obtained broadband SEDs for 105 bright
blazars ($F_{radio}>1$ Jy) and calculated $\nu_{peak}^{s}$ based on the simultaneous observations
of the Planck, Swift, and Fermi telescopes.
As a result, an average $\nu_{peak}^{s}$
of $\nu_{peak}^{s}$=$10^{13.1\pm0.1}$ Hz was determined
for FSRQs. For BL Lac blazars, this value is 
higher, and the distribution of $\nu_{peak}^{s}$ is
broader. The results of that work are well consistent
with \cite{2010ApJ...716...30A}, where empirical relations were derived
to determine $\nu_{peak}^{s}$ from the broadband spectral indices
(radio-optical and optical-X-ray) for 48 bright
blazars from the Fermi list. The values log~$\nu_{peak}^{s}\sim13$ and log~$\nu_{peak}^{s}\sim15$
were found for FSRQ and BL Lac blazars respectively.
In the second AGN catalogue of the
Fermi telescope (2LAC), an analytical formula similar to \cite{2010ApJ...716...30A} was
used for finding $\nu_{peak}^{s}$, and for most of the FSRQs
$\nu_{peak}^{s}$ < $10^{14}$ Hz was obtained, while for the BL Lac objects it was $\nu_{peak}^{s}$ > $10^{15}$ Hz \cite{2011ApJ...743..171A}.

Meyer et al. \cite{2011ApJ...740...98M} have measured $\nu_{peak}^{s}$ for a relatively large sample of 216 blazars.

Determination of the synchrotron peak frequency
($\nu_{peak}^{s}$), and the blazar type along with it, is an important
task for researchers of the AGN phenomenon,
as this parameter affects the distribution of radiating
particles by the energies in the jet, as well as physical
processes and the state of the matter in the emission
region. Making the measurements of $\nu_{peak}^{s}$
based on observations of a large number of blazars, one can
test other empirical relationships which are used to
compute this parameter in the lack of experimental data.

The aim of this work is to study the synchrotron
component of the spectral energy
distribution of a sample of 877 blazars.
The objects are systematically observed
with the RATAN-600 radio telescope.
361 sources of our sample are the BL Lac objects and candidates,
representing 25\% of all known blazars of this
type\footnote{according to the Roma-BZCAT, 5th edition},
 423 are the FSRQ-type blazars, and 93 are
the blazars of uncertain type. 

\section{The sample and observations}
We have studied 877 blazars from the RATAN-600 monitoring list,
 objects are presented in Table~\ref{table:list}.
The full version of Table~\ref{table:list} is available at the 
Strasbourg astronomical Data Center (CDS)\footnote{http://vizier.cfa.harvard.edu/}.
Notes for the columns in Table~\ref{table:list}: 
(1) NVSS name,
(2) alias,
(3) redshift $z$,
(4) R-band magnitude (USNO),
(5) the logarithm of the synchrotron peak frequency obtained in the present work,
(6) correlation coefficient between the experimental data and the theoretical curve used in
the calculations of $\nu_{peak}^{s}$,
(7) flux density at a frequency of 4.8 GHz
 and its standard error, obtained with the RATAN-600,
(8) blazar type based on the
position of the synchrotron component on the SED: LSP ISP HSP,
(9) blazar type according to the classification of \cite{2009A&A...495..691M},
(10) BL Lac type, based on the band in which it was originally identified: RBL --- radio-selected BL Lac,
	XBL --- X-ray-selected BL Lac.

\begin{small}
\begin{table*}[]
\setcaptionmargin{0mm}
\captionstyle{normal}
\onelinecaptionstrue
\caption{List of the studied blazars}
\label{table:list}
\medskip
\begin{tabular}{l|l@{\hspace{0.3cm}}|c|c|c|c@{\hspace{0.3cm}}|c@{\hspace{0.3cm}}|c|c|c}
\hline\hline
NVSS name       & Alias        &  $z$  & $R_{mag}$ & log $\nu_{peak}^{s}$ & $k$ & $F_{4.8GHz} \pm \sigma$, & SED   & Blazar     &  Selection \\
                &              &       &           &        [Hz]          &     &      Jy               & class & type       & type \\
\hline\hline
000520$+$052411 & BZQJ0005+0524           & 1.900 & 16.2 &  15.17 & 0.90 & $0.126\pm0.004$  & ISP & FSRQ            &  -    \\
000557$+$382015 & GB6B0003+3803           & 0.229 & 17.6 &  13.28 & 0.92 & $0.470\pm0.019$  & LSP & FSRQ            &  -    \\
000613$-$062335 & PKS0003-066             & 0.347 & 17.9 &  12.93 & 0.95 & $2.118\pm0.049$  & LSP & BL~Lac          &  RBL  \\
000649$+$242236 & CGRaBSJ0006+2422        & 1.684 & 18.8 &  14.27 & 0.91 & $0.132\pm0.012$  & LSP & FSRQ            &  -    \\
000759$+$471207 & BZBJ0007+4712           & 0.280 & 18.2 &  13.66 & 0.86 & $0.064\pm0.005$  & LSP & BL~Lac          &  RBL  \\
001031$+$105830 & PGC737                  & 0.089 & 15.8 &  13.92 & 0.94 & $0.120\pm0.005$  & LSP & FSRQ            &  -    \\
001101$-$261233 & PKS0008-264             & 1.096 & 18.8 &  13.97 & 0.91 & $0.600\pm0.024$  & LSP & FSRQ            &  -    \\
001354$-$042352 & PKS0011-046             & 1.075 & 19.7 &  12.58 & 0.92 & $0.230\pm0.009$  & LSP & FSRQ            &  -    \\
\hline\hline
\end{tabular}
\end{table*}
\end{small}

The redshifts of the objects are taken from the Roma-BZCAT catalogue\footnote{http://www.asdc.asi.it/bzcat/} \cite{2009A&A...495..691M} or from the NED.
The Roma-BZCAT is the most popular catalogue of the blazars, based on a large number of various surveys
with the use of observed data from different wavelength ranges. The average redshift for the FSRQs
is $z=1.446$, and for the BL Lac and candidates it is $z=0.443$.

Table~\ref{table:type} shows the distribution of objects by
their types. For the classification of blazars by the
optical spectrum, we used the Roma-BZCAT: 
FSRQ ---  flat-spectrum radio quasar,
BL~Lac --- BL Lacertae type object,
BL~Lac.cand. --- BL Lacertae candidate,
Blazar~un.type --- blazar of uncertain type.

We have made the division of BL Lac blazars (a
total of 454, including the candidates and blazars of
uncertain type) by the type of their detection (RBL,
XBL, or OBL) from literature data (see the references
in the BLcat catalogue\footnote{http://www.sao.ru/blcat/} \cite{2014A&A...572A..59M}).
The classification of
blazars by the SED type was done in this work
by the criteria set out in~\cite{1995PASP..107..803U}.

\begin{table}[]
\setcaptionmargin{0mm} \onelinecaptionsfalse \captionstyle{normal}
\caption{Sample subpopulation classes}
\label{table:type}
\medskip
\begin{tabular}{l|l|r}
\hline\hline
Designation criterion & Class & Number\\
\hline\hline
\multirow{4}{*}{Optical spectrum} & BL~Lac & 296 \\
		 & BL~Lac~cand. & 65 \\
		 & Blazar~un.type & 93\\
		 & FSRQ & 423\\
\hline
\multirow{3}{*}{Selection method} & RBL & 327\\
		       & XBL & 124 \\
		       & OBL & 3 \\
\hline
\multirow{3}{*}{SED type} & LSP & 611 \\
		 & ISP & 222 \\
		 & HSP & 42 \\
\hline
\hline
\end{tabular}
\end{table}

\begin{figure}[]
\setcaptionmargin{5mm}
\onelinecaptionsfalse
\captionstyle{normal}
\includegraphics[width=\textwidth]{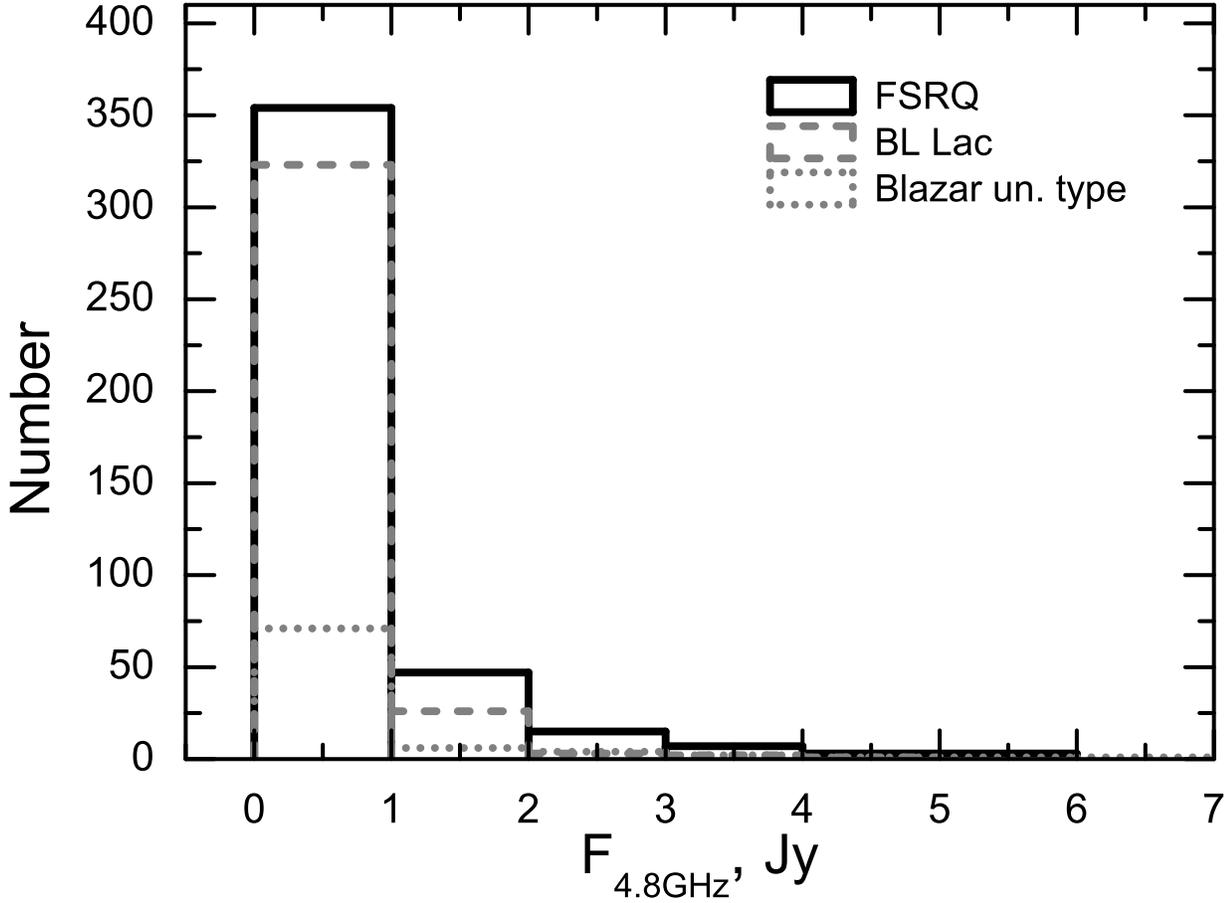} \\ 
\caption{The flux density distribution for the sources of the sample at 4.8 GHz, measured with the RATAN-600. 
Three objects (PKS\,B1226+023, PKS\,1253-055 и PKS\,1921-293) with the
$F_{4.8 GHz}>6$ Jy were excluded from distribution for clarity of the display}
\label{fluxes}
\end{figure}

The flux density distribution for the sources of the sample
at a frequency of 4.8 GHz is shown in Fig.~\ref{fluxes}.
The subsample of FSRQs consists of the brightest
objects at the radio band: the average flux
density at a frequency of 4.8 GHz (according to the
RATAN-600 data) for them is 0.736 Jy. The subsample
of BL Lac blazars is formed from the sources that
are weaker in the radio band with an average $F_{4.8 GHz}$=0.355 Jy.
Table~\ref{table:aver} lists the average values of some
parameters of the objects of our sample.

Most of the measurements of BL Lac objects are
available in the on-line BLcat catalogue. Some measurements
of FSRQs presented in this paper contain new
RATAN-600 observations carried out with two radiometric
complexes in 2014--2015. Parameters of the
antenna and the receiving systems of RATAN-600
secondary mirrors $\textnumero$1 and $\textnumero$2 are presented in Table~\ref{table:rad}
(the secondary mirrors are marked as ``1'' and ``2''
respectively). The columns contain: 
(1) central frequency in GHz,
(2) bandwidth in GHz,
(3) detection limit by the flux density per angular resolution unit (mJy/beam),
(4) angular resolution at the right ascension (arcsec) at the intermediate
antenna incidence angles ($\delta \sim 42^{\circ}$).

The technique of observations and calibration of
measurements are described, for example, in~\cite{2012A&A...544A..25M,2014A&A...572A..59M}.

\begin{table}[]
\setcaptionmargin{0mm} 
\captionstyle{normal}
\caption{The average values of some parameters for different subclasses of blazars
(the number of measurements is shown in the lower index)}
\label{table:aver}
\bigskip
\begin{tabular}{l|l|l|l|l}
\hline
\hline
Blazar     &  $z$ & log~$\nu_{peak}$ & $F_{4.8 GHz}$, & $R_{mag}$ \\
type       &      &   [Hz]           &   Jy        &           \\
\hline
\hline
FSRQ           & 1.446$_{422}$ & $13.4\pm1.0_{422}$ & 0.736$_{423}$ & 18.6$_{423}$ \\
BL~Lac         & 0.443$_{253}$ & $14.6\pm1.4_{360}$ & 0.355$_{359}$ & 17.3$_{358}$ \\
Blazar~un.type & 0.499$_{86}$  & $13.9\pm1.1_{93}$  & 0.739$_{93}$  & 17.2$_{92}$ \\
\hline
RBL            & 0.53$_{234}$ & $13.9\pm0.9_{326}$ & 0.624$_{329}$ & 14.3$_{325}$ \\
XBL            & 0.30$_{102}$ & $15.9\pm1.3_{124}$ & 0.089$_{123}$ & 16.7$_{123}$ \\
\hline
\hline
\end{tabular}
\end{table}

\begin{table}[]
\setcaptionmargin{0mm}
\captionstyle{normal}
\onelinecaptionsfalse
\caption{Some parameters of the RATAN-600 antenna and continuum radiometers}
\label{table:rad}
\medskip
\begin{tabular}{cc|cc|cc|c}
\hline\hline
\multicolumn{2}{c|}{$f_{0}$,}  & \multicolumn{2}{c|}{$\Delta$$f_{0}$,} & \multicolumn{2}{c|}{$\Delta$$F$,}  & $\theta_{RA},$    \\
\multicolumn{2}{c|}{GHz}    & \multicolumn{2}{c|}{GHz}           & \multicolumn{2}{c|}{mJy/beam}   & arcsec\\
\hline
      1     &     2           &    1        &     2            &    1     &     2     &         \\
\hline
\hline
$21.7$ & $21.7$ & $2.5$  & $2.5$  &  $70$  & $88$  &   11 \\
$11.2$ & $11.2$ & $1.4$  & $1.0$  &  $20$  & $20$  &   16 \\
$7.7$  & -      & $1.0$  & -      &  $25$  & -     &   22 \\
$4.8$  & $4.8$  & $0.9$  & $0.8$  &  $8$   & $11$  &   36 \\
$2.3$  & -      & $0.4$  & -      &  $30$  & -     &   80 \\
$1.1$  & -      & $0.12$ & -      &  $160$ & -     &  170 \\
\hline
\hline
\multicolumn{7}{{p{.4\textwidth}}}{{\footnotesize 
Column description:
(1)~-- central frequency	,
(2)~-- bandwidth,
(3)~-- flux density detection limit per beam,
(4)~-- angular resolution (at RA).}} \\
\hline
\hline
\end{tabular}
\end{table}

\section{The synchrotron peak frequency estimation}

The SEDs were constructed in the log~$\nu$ -- log~$\nu F_{\nu}$ plane.
 The synchrotron component can be described by a polynomial of the second or
third degree:
\[log~(\nu F_{\nu})=A(log\nu)^2+B(log\nu)+C\]
\[log~(\nu F_{\nu})=A(log\nu)^3+B(log\nu)^2+C(log\nu)+D,\]
where A, B, C and D are the coefficients.
Hence:
\[log~\nu_{peak}=-B/2A.\]

We used the ASDC SED Builder Tool\footnote{http://tools.asdc.asi.it}
\cite{2011arXiv1103.0749S} to calculate the synchrotron peak frequency.
It allows building broadband SEDs of sources and
approximate the experimental data by a theoretical
curve. The system is based on the local catalogues covering
a wide range of the electromagnetic spectrum,
from radio to gamma-rays.

\begin{figure*}[]
\setcaptionmargin{5mm}
\onelinecaptionsfalse
\includegraphics[width=\textwidth]{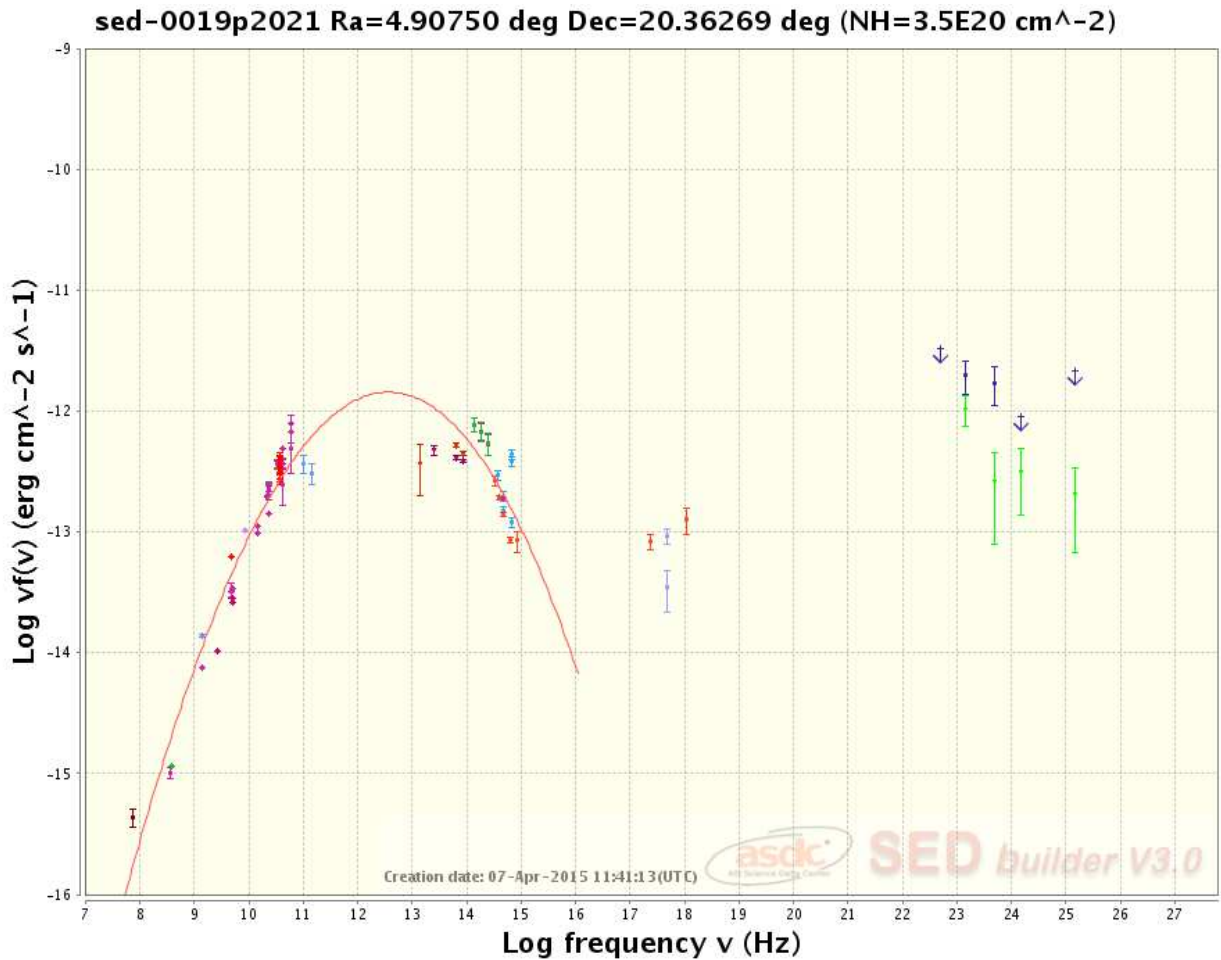} \\ 
\captionstyle{normal}
\caption{The SED of the PKS\,0017$+$200, visualised with the ASDC SED Builder Tool}
\label{SED}
\end{figure*}

In the present study we used the polynomial of the
second or third degree. The correlation coefficient is
given in the sixth column of Table~\ref{table:list}.
The example of the approximation for the PKS\,0017$+$200 given in Fig.~\ref{SED}.

Determination of the synchrotron peak frequency
often depends on the data set and calculation method,
because of the inhomogeneity of the experimental data.
If there is a lot of measurements at one band but only a few points
at the other bands, even overall measurements number is big,
there will be a weak correlation between 
experimental data points and theoretical curve,
and, therefore, uncertain estimation of the $\nu_{peak}^{s}$ parameter.

\begin{figure*}[]
\setcaptionmargin{5mm}
\onelinecaptionsfalse
\includegraphics[width=\textwidth]{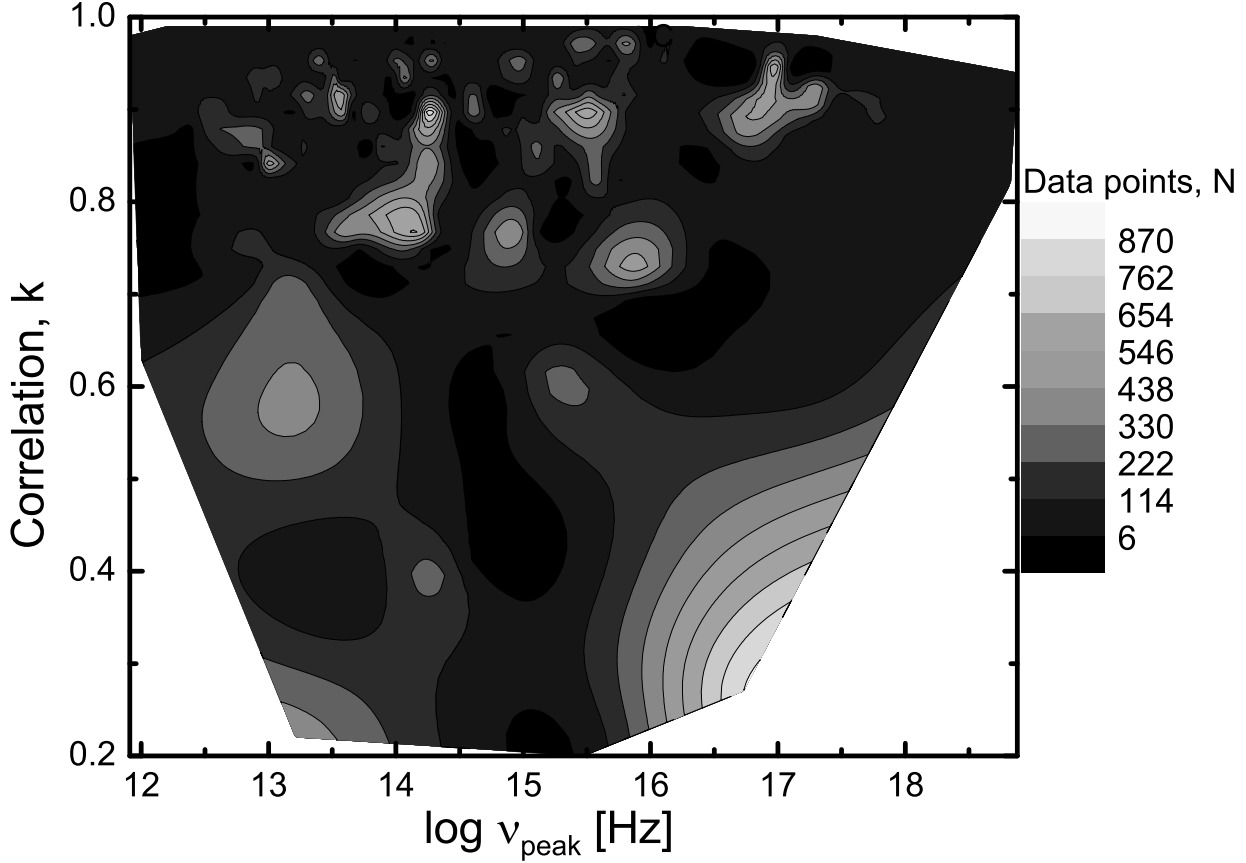} \\ 
\captionstyle{normal}
\caption{The $\nu_{peak}^{s}$---correlation coefficient---measurements number relation for BL Lacs}
\label{korr}
\end{figure*}

Fig.~\ref{korr} gives an example
of the relationship between the calculated value of
$\nu_{peak}^{s}$, the correlation coefficient, and the number of
measurements for BL Lac blazars of the sample. The
colours show the number of measurements N used for
construction of the SED curve for each object: white
corresponds to the maximum number of measurements
(N = 870), and black means the minimum
(N = 6). It is clearly seen that a low correlation of
the theoretical curve and the experimental data is
observed both for a small and a large number of
measurements (700--900 points). This is related not
only with inhomogeneity of measurements at certain
frequencies but also with object variability, as a result
of which a large scatter of data is observed when
non-simultaneous measurements are used. And conversely,
at a small number of measurements (up to
one hundred), high correlation can be observed, when
several points are easily described by any polynomial.

Fig.~\ref{korr} also demonstrates that the region of
$10^{17} < \nu_{peak}^{s} < 10^{19}$ Hz differs by a small number
of measurements (in most cases up to one hundred).
Therefore, the $\nu_{peak}^{s}$ values obtained for the HSP
blazars can be refined and calculated more reliably
with the increasing number of measurements at these frequencies.

An overestimation of $\nu_{peak}^{s}$
is possible when taking into account
thermal radiation in the optical/UV bands;
 in some objects this thermal component makes
a significant contribution~\cite{2012MNRAS.420.2899G}.
The deficiency of observational data in the X-rays, on the contrary, leads
to an underestimation of $\nu_{peak}^{s}$.

\section{Results}
\subsection{The $\nu_{peak}^{s}$ values}

\begin{figure*}[]
\setcaptionmargin{5mm}
\onelinecaptionstrue
\includegraphics[width=\textwidth]{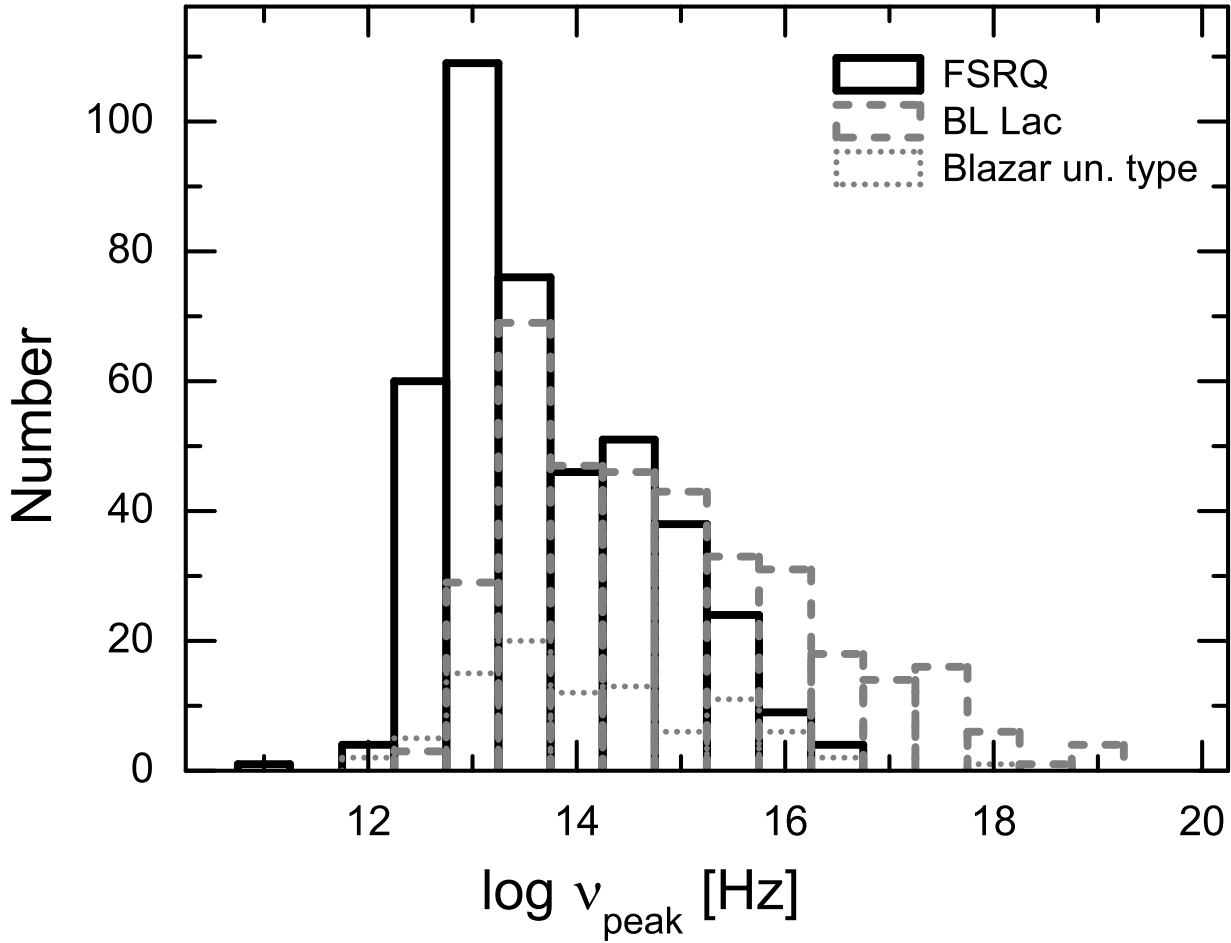} \\ 
\captionstyle{normal}
\caption{Distribution of $\nu_{peak}^{s}$ for FSRQs, BL Lacs and blazars of uncertain type}
\label{Hist}
\end{figure*}

The calculations of $\nu_{peak}^{s}$ were made in the observers frame.
The $\nu_{peak}^{s}$ have been obtained for 875 blazars of the sample
and are presented in the fifth column of Table~\ref{table:list}
We could not estimate $\nu_{peak}^{s}$ 
for two objects TEX\,0537$+$251 and BZQ\,J1102$+$5941
because lack of measurements at frequencies exceeding $10^{15}$ Hz.
The distribution of $\nu_{peak}^{s}$ for FSRQs and BL Lacs 
is presented in Fig.~\ref{Hist}.

The average values of the $z$, log~$\nu_{peak}^{s}$, $F_{4.8GHz}$ and $R_{mag}$
are given in Table~\ref{table:aver}.

The $\nu_{peak}^{s}$ and $F_{4.8~GHz}$ samples for FSRQs, BL Lacs and RBLs, XBLs are drawn from different distributions,
according to the Kolmogorov-Smirnov (K-S) test (at the 0.05 significance level).

\begin{figure*}[]
\setcaptionmargin{5mm}
\onelinecaptionsfalse
\includegraphics[width=\textwidth]{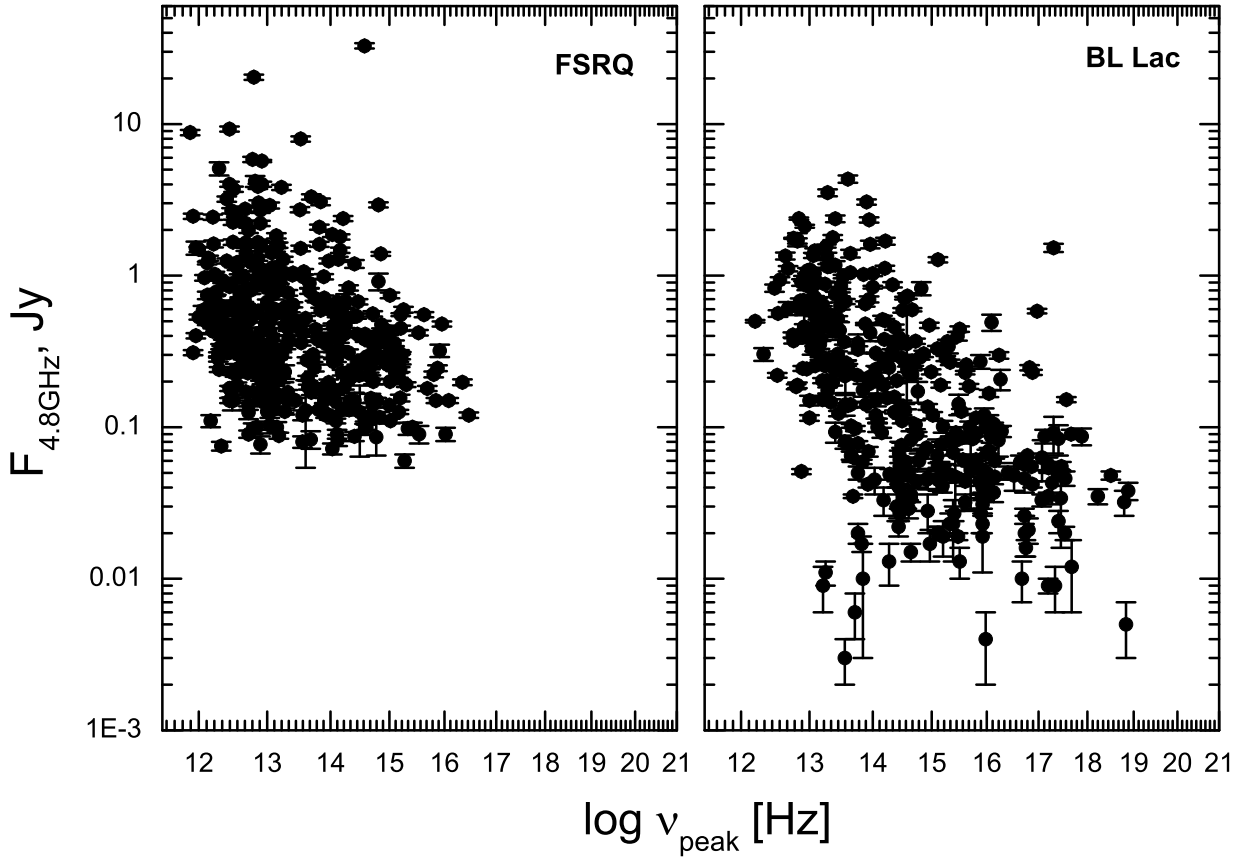} \\ 
\captionstyle{normal}
\caption{The $\nu_{peak}^{s}$---flux density at 4.8 GHz ($F_{4.8GHz}$) relation
for FSRQs and BL Lacs. The mean value of $\nu_{peak}^{s}$ is
$10^{13.4\pm1.0}$ Hz for FSRQs and $10^{14.6\pm1.4}$ Hz for BL~Lacs and candidates}
\label{FSRQ}
\end{figure*}

The $\nu_{peak}^{s} - F_{4.8GHz}$ relations for the FSRQs and BL Lacs 
are shown in Fig.~\ref{FSRQ}.
The flux density of blazars obtained with the RATAN-600 at a frequency
of 4.8 GHz is presented in Table~\ref{table:list} (column 7).
The distribution of $\nu_{peak}^{s}$ for the BL Lac objects and
candidates is wider, with an average of $10^{14.6\pm1.4}$ Hz ($10^{11.9}$--$10^{18.9}$ Hz).
The distribution of $\nu_{peak}^{s}$ for the
FSRQs has an average value of $10^{13.4\pm1.0}$ Hz, and
most of the values are located in the range from $10^{11.9}$ to $10^{16.5}$ Hz.

\subsection{RBL and XBL objects}

\begin{figure*}[]
\setcaptionmargin{5mm}
\onelinecaptionsfalse
\includegraphics[width=\textwidth]{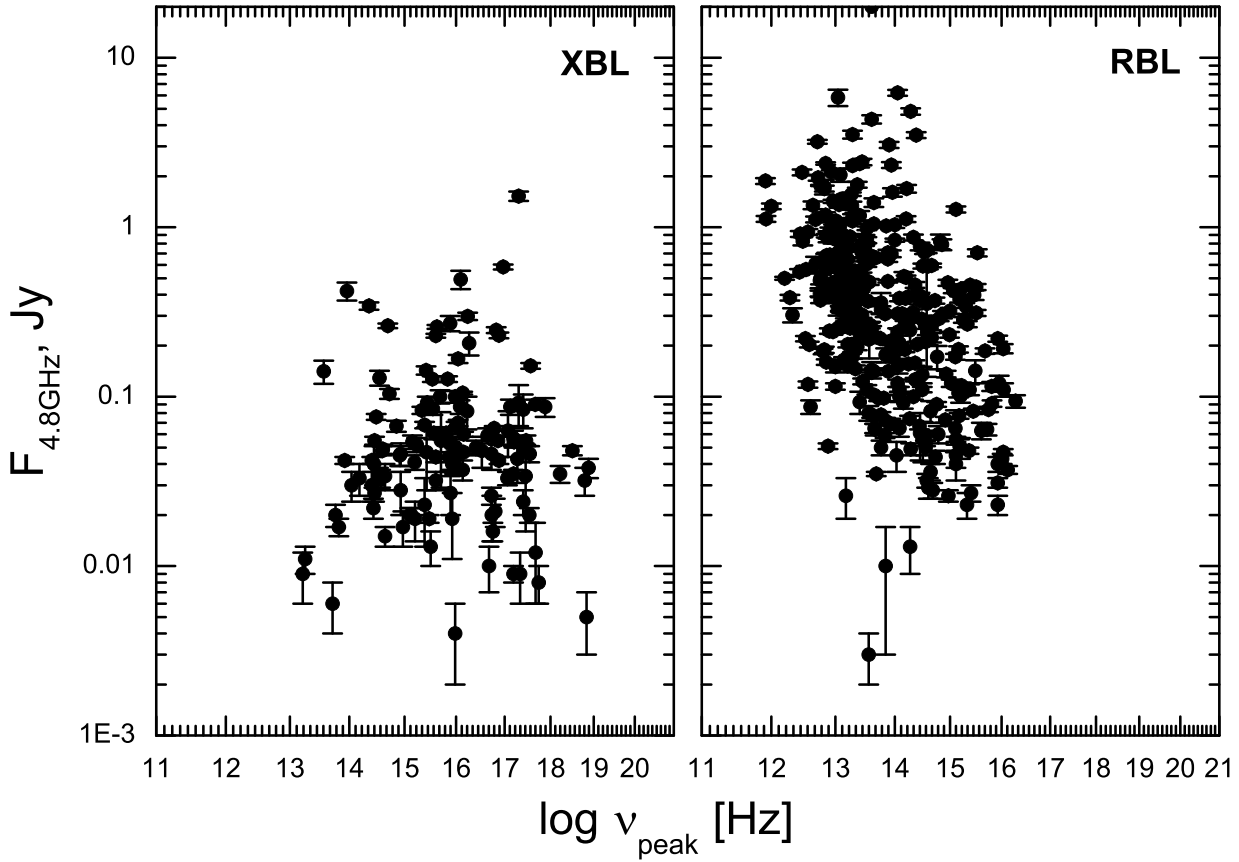} \\ 
\captionstyle{normal}
\caption{The $\nu_{peak}^{s}$---$F_{4.8GHz}$ relation for RBL and XBL objects.
The mean value of $\nu_{peak}^{s}$ is $10^{13.9\pm0.9}$ and $10^{15.9\pm1.3}$ Hz for RBLs and XBLs, respectively}
\label{rbl}
\end{figure*}

The classification of BL Lac objects to the RBL
and XBL types was made using literature data
and is presented in Table~\ref{table:list} (column 10).
The sample consist of 327 RBLs, 124 XBLs and 3 OBLs.
The $\nu_{peak}^{s} - F_{4.8GHz}$ relations for the RBL and XBL objects
are shown in Fig.~\ref{rbl}.
On average, $\nu_{peak}^{s}$ values for the RBLs is less than that for the XBLs.
The average $\nu_{peak}^{s}$ for the RBLs is  $10^{13.9\pm0.9}$ Hz,
and for the XBLs it is $10^{15.9\pm1.3}$ Hz.

The distribution of peak frequencies for the XBLs
is broader ($10^{13.2}$ -- $10^{18.9}$ Hz) than for the RBLs
($10^{11.9}$ -- $10^{16.3}$ Hz). The average flux densities $F_{4.8GHz}$ 
for the XBLs and RBLs differ significantly (see Table~\ref{table:aver}).

\subsection{UHBL candidates}

Table~\ref{table:uhbl} gives a list of blazars which
were considered in \cite{2006A&A...445..441N} as UHBL candidates (ultra-high energy-peaked BL Lac objects).
The blazars with a synchrotron component peak frequency
of $\nu_{peak}^{s}$~>~$10^{19}$ Hz were related to this class
of objects. The first column gives the name of an
object. The second and third columns of Table~\ref{table:uhbl}
represent the measurements of log~$\nu_{peak}^{s}$
in \cite{2006A&A...445..441N} and in the present work, respectively. The
fourth column lists the flux density at a frequency
of 4.8 GHz and its error, obtained with the RATAN-600.
The fifth column contains the amplitude of object
variability (in percent) at a frequency of 4.8 GHz and
the number of measurements, which are indicated
at the lower index, according to the RATAN-600
data. The amplitude variability was determined by the formula:
\[Var_{F}=\frac{(F_{i}-\sigma_{i})_{max}-(F_{i}+\sigma_{i})_{min}}
{(F_{i}-\sigma_{i})_{max}+(F_{i}+\sigma_{i})_{min}},\]
where $F_{max}$ and $F_{min}$ are the maximum and minimum flux density,
$\sigma_{F_{max}}$ and $\sigma_{F_{min}}$ are their errors.
The sixth column presents the type of a BL Lac by the
identification band, taken from the literature.
No UHBL candidates were found in our sample
according to this classification.
The maximum value of $\nu_{peak}^{s}$ in this sample is measured
for the objects 1ES\,0229+200 ($10^{18.5}$~Hz), 1ES\,0502+675 ($10^{18.9}$~Hz) and
RXS\,J1458.4+4832 ($10^{18.8}$~Hz).
All objects listed in
Table~\ref{table:uhbl} are belong to the XBL subclass of BL Lac objects.
The UHBL candidates list is composed of the
objects faint in the radio domain, with the flux densities not exceeding 100 mJy at a frequency
of 4.8 GHz (column 4 of Table~\ref{table:uhbl}).
Almost all the objects are variable in the radio band: the amplitude
of variability ranges from a few to tens of percent.
Some of them were observed insufficiently, only
3--5 times. In other frequency ranges they have also
been scarcely measured. Being historically discovered
in X-rays, they were hence probably mostly observed
in this very spectral region and thus became UHBL candidates.

\begin{table*}[h]
\setcaptionmargin{0mm}
\onelinecaptionsfalse
\captionstyle{flushleft}
\caption{UHBL candidates (log~$\nu_{peak}^{s} > 19$) from \cite{2006A&A...445..441N}.
The log~$\nu_{peak}^{s}$ values are presented for 2006 in column 2 (from \cite{2006A&A...445..441N}) and for 2015 in column 3 (results of this work, signed RATAN). 
The flux density and variability index values, measured at 4.8 GHz with the RATAN-600, are shown in the columns (4) and (5)}
\label{table:uhbl}
\bigskip
\begin{tabular}{c|c|c|c|c|c}
\hline
 Name & log~$\nu_{peak}^{s}$, & log~$\nu_{peak}^{s}$, & $F_{4.8GHz}$, &  Var\,$F_{4.8_{Nobs}}$, & Selected type \\
      &  {\cite{2006A&A...445..441N}}   &    RATAN      &      Jy    &  \% & \\
\hline
1ES\,0229+200     & 19.45 & 18.50 & $0.049\pm0.003$ & $7.5_{5}$    & XBL \\
RXS\,J0314.3+0620 & 19.57 & 16.13 & $0.038\pm0.005$ & $1.5_{5}$    & XBL \\
2E\,0323+0214     & 19.87 & 15.92 & $0.041\pm0.003$ & $17.1_{7}$   & XBL \\
2E\,0414+0057     & 20.71 & 16.78 & $0.065\pm0.002$ & $22.0_{7}$   & XBL \\
1ES\,0502+675     & 19.18 & 18.88 & $0.038\pm0.005$ & $18.4_{6}$   & XBL \\
EXO\,0706.1+5913  & 21.05 & 17.88 & $0.087\pm0.011$ &  $1.8_{6}$   & XBL \\
RXS\,J0847.2+1133 & 19.13 & 17.40 & $0.024\pm0.004$ &    -         & XBL \\
1ES\,0927+500     & 21.13 & 17.40 & $0.084\pm0.019$ & $56.6_{5}$   & XBL \\
RXS\,J1008.1+4705 & 19.67 & 17.33 & $0.009\pm0.003$ & $76.5_{3}$   & XBL \\
RXS\,J1012.7+4229 & 20.97 & 17.46 & $0.055\pm0.004$ & $13.8_{8}$   & XBL \\
EXO\,1149.9+2455  & 19.83 & 16.72 & $0.026\pm0.003$ & $12.2_{3}$   & XBL \\
PG\,1218+304      & 19.14 & 16.86 & $0.055\pm0.003$ & $8.3_{5}$    & XBL \\
RXS\,J1319.5+1405 & 19.67 & 15.15 & $0.055\pm0.005$ & $12.4_{5}$   & XBL \\
RXS\,J1341.0+3959 & 20.97 & 14.59 & $0.048\pm0.003$ & $8.2_{6}$    & XBL \\
RXS\,J1353.4+5601 & 19.67 & 15.92 & $0.019\pm0.008$ &  -           & XBL \\
RXS\,J1410.5+6100 & 20.97 & 14.44 & $0.040\pm0.011$ &  -           & XBL \\
2E\,1415+2557     & 19.24 & 17.54 & $0.046\pm0.005$ &  -           & XBL \\
RXS\,J1456.0+5048 & 19.94 & 16.22 & $0.082\pm0.018$ & $7.6_{5}$    & XBL \\
RXS\,J1458.4+4832 & 21.46 & 18.83 & $0.005\pm0.002$ & -            & XBL \\
1ES\,1533+535     & 19.68 & 16.72 & $0.047\pm0.009$ & $14.6_{5}$   & XBL \\
RXS\,J1756.2+5522 & 19.90 & 17.27 & $0.043\pm0.012$ & $19.1_{5}$   & XBL \\
RXS\,J2304.6+3705 & 21.01 & 17.53 & $0.020\pm0.003$ & $9.5_{8}$    & XBL \\
\hline
\end{tabular}
\end{table*}

\subsection{VLSP candidates}

Along with the high/low-synchrotron peaked blazars,
there is so-called very low synchrotron peaked blazars (VLSP, $\nu_{peak}^{s} < 10^{13}$ Hz)
were recently denoted~\cite{2005MNRAS.356..225A,2010A&A...512A..74M}.
The maximum of the synchrotron component in them falls into the IR/mm spectral region.

Ghisellini et al. \cite{1998MNRAS.301..451G} described the relationship between
the maximum energy of electrons
$\gamma_{peak}$ and the density of the total energy $(U_{ph}+U_{B})$,
where $U_{ph}$ is the density of photon energy, and $U_{B}$ is the energy
density of the magnetic field in the jet:
\[\gamma_{peak}\sim(U_{ph}+U_{B})^{-0.6}.\]
If we assume that the total energy density is $U=L/R^{2}$, where $L$ is the luminosity of the jet and
$R$ is its size, and that the synchrotron peak frequency $\nu_{peak}^{s} \propto \gamma_{peak}^2$,
then objects of high luminosity have a lower peak frequency. Hence, objects with a very
low value of $\nu_{peak}^{s}$ would often be quite bright radio
sources. In fact, most of the objects in our sample
of VLSP candidates are the FSRQs (41\% of the
total number of FSRQs) and only 9\% are the BL Lacs.

\section{Discussion}
The results of our work are consistent with the results of other authors obtained from samples
with a significantly smaller number of objects, such as:

-- study of the synchrotron component
of 300 BL Lac objects in~\cite{2006A&A...445..441N} based on non-simultaneous
literature data; for the majority of objects, $\nu_{peak}^{s}$ is within $10^{13-14}$ Hz.

-- study of the synchrotron component of 105 bright blazars based on the simultaneous
measurements of the Planck, Swift, and Fermi
telescopes \cite{2012A&A...541A.160G},
the average $\nu_{peak}^{s}$ for FSRQs is $10^{13.1\pm0.1}$ Hz.

-- $\nu_{peak}^{s}$ values estimation using broadband spectral indices
 $\alpha_{ro}$ and $\alpha_{ox}$
(between the frequencies of  5~GHz, 5000~\AA~and 1~keV) was made in  
\cite{2010ApJ...716...30A,2011ApJ...743..171A}; as a result
 $\nu_{peak}^{s}$=$10^{13.02\pm0.35}$ Hz for FSRQs and
a broad distribution for BL~Lasc --- from
the lowest to the highest frequencies.

In this study most of the objects have $\nu_{peak}^{s}$ $10^{13-14}$ Hz.
The HSP blazars are quite
rare in the sample --- 5\% of the total number, these are
mostly the BL Lac objects.

The extreme $\nu_{peak}^{s}$ values (greater than $10^{19}$ Hz)
were not confirmed for 22 objects from the list of
UHBL candidates from \cite{2006A&A...445..441N}. It is easy to notice that in
the presence of few points on the SED, $\nu_{peak}^{s}$
can often be overestimated, especially if a BL Lac was
detected in the high-frequency spectral range. This is
true for our sample: all the UHBL candidates are the XBL blazars.

The distributions of $\nu_{peak}^{s}$ for FSRQs and BL Lac
objects have different characters, which is sometimes
interpreted within different morphology of the objects
and possible evolution of FSRQs into BL Lac objects
\cite{1995PASP..107..803U}.

\section{Summary}

We have studied the synchrotron component of the spectral energy distribution (SED)
on the sample of 877 blazars of various subclasses.
We made an estimation of the synchrotron peak frequency $\nu_{peak}^{s}$,
using Roma-BZCAT catalogue and ASDC SED Builder Tool.
The following results were obtained:
\begin{itemize}
\item The $\nu_{peak}^{s}$ values were found for 875 objects.
We further classified them by the SED type:
611 (70\%) LSPs, 222 (25\%) --- ISPs and 42 (5\%) --- HSPs.
In the cases of lack of measurements at different wavebands or variable object,
$\nu_{peak}^{s}$ value estimation is strongly depend on the choice of observational data (active or quiescent state),
range of the bands and polynomial degree used for the approximation. 

\item The distribution of $\nu_{peak}^{s}$
peak smoothly decreases towards
higher frequencies (Fig.~\ref{Hist}). Only 5\% of the sample are
HSP blazars , most of them are the BL Lacs. It is possible that
objects in which electrons are accelerated to very
high energies are rare, or it is result of the selection effect

\item The distribution of $\nu_{peak}^{s}$ is different for FSRQs and BL~Lacs.
For the BL Lac objects and candidates it is broader and
shifted to the more high-frequency region,
the average $\nu_{peak}^{s}$ value is $10^{14.6\pm1.4}$ Hz.
 For the FSRQs the average $\nu_{peak}^{s}$ value is $10^{13.4\pm1.0}$ Hz.
 Statistical tests have shown that $\nu_{peak}^{s}$ and
 flux density $F_{4.8GHz}$ form different distributions
 for FSRQs and BL Lacs.

\item The $\nu_{peak}^{s}$ and $F_{4.8GHz}$
are also form different distributions
for RBL and XBL blazars (at 0.05 confidence level).
The average $\nu_{peak}^{s}$ is $10^{13.9\pm0.9}$ Hz and $10^{15.9\pm1.3}$ Hz
for RBLs and XBLs, respectively.

\item We obtained new estimates of $\nu_{peak}^{s}$ for 22 UHBL candidates
and found that none of them have 
$\nu_{peak}^{s}$ > $10^{19}$ Hz. The fact that
the majority of measurements were performed in
the X-ray band, where they all were detected, was
probably the reason of large $\nu_{peak}^{s}$
 values in \cite{2006A&A...445..441N},
where these blazars were classified as Ultra-high energy-peaked.
Most of these objects belong to the HSP and partly to the ISP
blazars, according to the data collected in
the catalogue of Massaro et al.\cite{2009A&A...495..691M}.

\item We found that 41\% of FSRQs and only 9\% of BL Lacs in our sample
could be considered as candidates to the very low synchrotron peaked blazars
(with $\nu_{peak}^{s} < 10^{13}$ Hz).
\end{itemize}

\begin{acknowledgments}
The RATAN-600 observations were carried out with the
financial support of the Ministry of Education and Science of the Russian
Federation and partly with the support of the Russian Foundation for Basic Research (project \textnumero 12-02-31649).
The authors acknowledges support through the Russian Government Program of Competitive Growth of the Kazan Federal University.

We used on-line version of the Roma-BZCAT catalogue and the SED Builder Tool at the ASI Science Data Center (ASDC) website,
therefore we are grateful to the ASDC staff.
This research has made use of the NASA/IPAC Extragalactic Database (NED) which is operated by the Jet Propulsion Laboratory, California Institute of Technology, under contract with the National Aeronautics and Space Administration.
\end{acknowledgments}

\bibliographystyle{AstroBull}
\bibliography{mingaliev}

\end{document}

%% file: sao_cmd_author.tex
\def\squareforqed{\hbox{\rlap{$\sqcap$}$\sqcup$}}

\def\sq{\ifmmode\squareforqed\else{\unskip\nobreak\hfil
\penalty50\hskip1em\null\nobreak\hfil\squareforqed
\parfillskip=0pt\finalhyphendemerits=0\endgraf}\fi}

\def\utw{\smash{\rlap{\lower5pt\hbox{$\sim$}}}}

\def\udtw{\smash{\rlap{\lower6pt\hbox{$\approx$}}}}

\def\diameter{{\ifmmode\mathchoice
{\ooalign{\hfil\hbox{$\displaystyle/$}\hfil\crcr
{\hbox{$\displaystyle\mathchar"20D$}}}}
{\ooalign{\hfil\hbox{$\textstyle/$}\hfil\crcr
{\hbox{$\textstyle\mathchar"20D$}}}}
{\ooalign{\hfil\hbox{$\scriptstyle/$}\hfil\crcr
{\hbox{$\scriptstyle\mathchar"20D$}}}}
{\ooalign{\hfil\hbox{$\scriptscriptstyle/$}\hfil\crcr
{\hbox{$\scriptscriptstyle\mathchar"20D$}}}}
\else{\ooalign{\hfil/\hfil\crcr\mathhexbox20D}}%
\fi}}

